\documentclass[pra,reprint,english,showpacs]{revtex4-1}
\usepackage[T1]{fontenc}
\usepackage[latin9]{inputenc}
\usepackage{array}
\usepackage{float}
\usepackage{amsmath}
\usepackage{graphicx}

\usepackage{babel}

\begin{document}
\newcommand{\ket}[1]{|#1\rangle}
\newcommand{\bra}[1]{\langle#1|}

\title{How to measure diffusional decoherence in multimode rubidium vapor
memories? }

\author{Rados{\l{}}aw Chrapkiewicz} 
\email{radekch@fuw.edu.pl}
\affiliation{Insitute of Experimental Physics, University of Warsaw, ul. Ho\.z{}a 69, 00-681, Warsaw, Poland}

\author{Wojciech Wasilewski}
\affiliation{Insitute of Experimental Physics, University of Warsaw, ul. Ho\.z{}a 69, 00-681, Warsaw, Poland}

\author{Czes{\l{}}aw Radzewicz}
\affiliation{Insitute of Experimental Physics, University of Warsaw, ul. Ho\.z{}a 69, 00-681, Warsaw, Poland}


\date{\today}

\begin{abstract}
Diffusion is the main limitation of storage time in spatially multimode
applications of warm atomic vapors. Precise knowledge of diffusional
decoherence in the system is desired for designing most of vapor memory
setups. Here we present a novel, efficient and direct method of measuring
unbiased diffusional decoherence, clearly distinguished from all other
decoherence sources. We found the normalized diffusion coefficients
of rubidium atoms in noble gases to be as follows: neon 0.20 cm$^{2}$/s,
krypton 0.068 cm$^{2}$/s and we are the first to give an experimental
result for rubidium in xenon: 0.057 cm$^{2}$/s. Our method consists
in creating, storing and retrieving spatially-varying atomic coherence.
Raman scattering provides a necessary interface to the atoms that
allows for probing many spatial periodicities of atomic coherence
concurrently. As opposed to previous experiments the method can be
used for any single sealed glass cell and it does not require any
setup alterations during the measurements and therefore it is robust
and repeatable. 
\end{abstract}
\maketitle

\section{Introduction}

\label{intro}

In recent years warm atomic ensembles have been widely used in many
applications in quantum optics and atomic physics. The most promising
applications include quantum repeaters \cite{Duan2001b}, quantum
memories \cite{Appel2008} and ultraprecise magnetometry \cite{Chalupczak2012}.
They have also been shown to be an effective medium for four-wave
mixing processes \cite{McCormick2007}, electromagnetically induced
transparency (EIT) \cite{Fleischhauer2005} and slow light generation
\cite{Matsko2001}.

An advantage of using warm atomic gas is undoubtedly simplicity of
performing experiments and obtaining large optical depths. However,
it entails fast thermal motion of atoms. Atomic motion limits the
efficiency of many modern systems that use multimode properties of
atomic ensembles. At best this motion can be slowed down and made
diffusive by addition of a suitable buffer gas. Since diffusion is
usually the main source of decoherence in multimode vapor memory systems,
ability to measure and control its speed is highly desired. In particular
it would be very useful to distinguish pure diffusional decoherence
from other decoherence effects.

There is a number of currently developing systems where diffusion
is the main limitation. An important example of such a system is storing
and retrieving transverse modes and images in gradient echo memory
\cite{Glorieux2012,Higginbottom2012,Clark2013,Luo2013}, in collective
Raman scattering \cite{Chrapkiewicz2012} or in EIT \cite{Firstenberg2012}.
Typically diffusional motion of atoms in a buffer gas limits the storage
time \cite{Shuker2008,Vudyasetu2008}, restricts the number of spatial
modes retrieved \cite{Chrapkiewicz2012} or broadens the EIT spectrum
\cite{Shuker2007}. Knowledge of the exact diffusion coefficient is
particularly important for designing experiments with diffraction
cancellation \cite{Firstenberg2008,Firstenberg2009,Firstenberg2012,Yankelev2013}.
Further progress would be significantly facilitated if one possessed
convenient, robust and repeatable method for precise diagnostic of
decoherence in the actual cell of a particular setup.

However, the available methods of measuring diffusion decoherence
are indirect and require either variation of buffer gas pressure,
prior knowledge of other sources of decoherence \cite{Franzen1959}
or setup alterations within a single measurement \cite{Bicchi1980,Glassner1996}.
This makes them unsuitable for modern experiments where exact knowledge
of diffusional decoherence of a single sealed glass cell in a specific
setup is required.

Here we propose a novel, direct method which allows us to measure
the diffusion in any given cell. The measurement provides more than
enough data to verify its self-consistency and single out the diffusion
from other motion-independent sources of decoherence. The method should
be relatively easy to incorporate into a number of quantum memory
setups.

As a demonstration we measure the diffusion coefficients of rubidium
in neon, krypton and xenon in sealed glass cells at a pressure of
a few torrs. These results will be useful for designing future experiments,
since the data available till now is rather scarce and, most importantly,
it was retrieved using indirect methods \cite{Franzen1959,Bernheim1962,McNeal1962,Arditi1964,Franz1965,Gozzini1967,Bouchiat1972,Vanier1974,Bicchi1980}.
The data available for neon is inconsistent and that for krypton is
hardly available \cite{Bouchiat1972,Higginbottom2012}.

We also recommend using xenon as a buffer gas, for which we provide
the very first experimental data as far as we know. Despite the latest
applications of hyperpolarized xenon \cite{Fink2005}, the diffusion
coefficient of rubidium in this gas has only been deduced from cross
sections of velocity changing collisions \cite{Gibble1991} or interaction
potentials \cite{Hamel1986}.

This paper is organized as follows: in Sec. 2 we introduce the principles
of our method, Sec. 3 describes in detail the experimental implementation,
Sec. 4 contains the experimental results together with the reference
data available. Finally, Sec. 5 concludes the paper.

\section{Method}

\begin{figure}
\includegraphics[width=8cm]{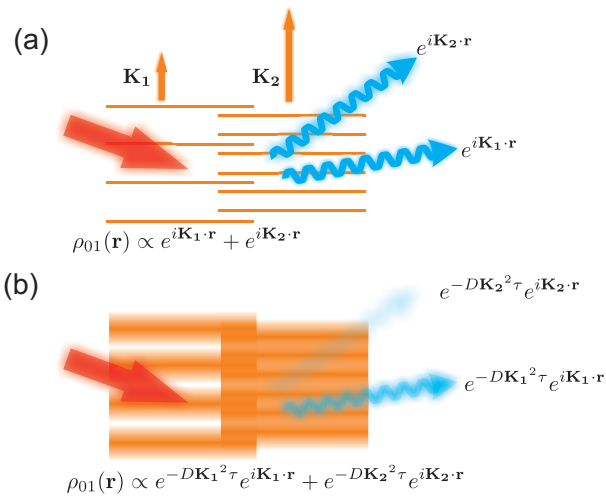} \centering \caption{(Color online) (a) Planewave components of spatially dependent atomic
coherence $\rho_{01}(\mathbf{r})$ act as diffraction gratings deflecting
laser beam at different angles. (b) Components of high periodicity
are blurred faster due to diffusion, therefore the intensity of deflected
light will decay faster for higher angles of deflection. During storage
time $\tau$, the pattern component of specific periodicity corresponding
to the wave vector $\mathbf{K}$ blurs with decay rate $D\mathbf{K}^{2}$,
where $D$ is the diffusion coefficient. \label{fig:Blurr}}
\end{figure}

\subsection{General idea}

Quantification of the diffusive motion of atoms using optical methods
can be done in three general steps. At first a group of atoms has
to be chosen and distinguished from others by changing their internal
state. Then the atoms are let to spread due to diffusion in the absence
of light. In the third stage one probes the group and observes the
effects of the spread. This general scheme has various implementations
\cite{Franzen1959,Bernheim1962,McNeal1962,Arditi1964,Franz1965,Gozzini1967,Bouchiat1972,Vanier1974,Bicchi1980}
which typically consisted in exciting and probing a pencil-shaped
atomic group using light beams. Instead here we create and, after
a certain delay, probe patterns of spatially dependent atomic coherence
$\rho_{01}(\mathbf{r})$ between two long lived atomic levels $\ket0$
and $\ket1$.

Patterns of atomic coherence are created through spontaneous (Stokes)
Raman scattering. Each pattern comprises many plane-wave components
with different periodicities. Those components decay at different
rates due to diffusive motion of the atoms. After a certain storage
time relative contribution of each plane-wave component can be measured
by driving the anti-Stokes scattering. Then each plane-wave component
acts as a diffraction grating deflecting driving laser beam as illustrated
in Fig. \ref{fig:Blurr} (a). By measuring the intensity of the anti-Stokes
scattering light as a function of deflection angle and time between
pattern creation and readout, we can calculate the decay rate of different
plane-wave components constituting atomic coherence. We rely on the
fact that in the diffusion process periodic patterns of atomic coherence
does not change their period but they are blurred over time as depicted
in Fig. \ref{fig:Blurr} (b). As atoms move, coherence at a specific
point $\mathbf{r}_{0}$ will reshuffle its values with the neighboring
points. Evolution of atomic coherence in the dark will be described
by the equation of diffusion with a coefficient $D$ and homogeneous
depolarization with a rate $\gamma_{0}$ \cite{Lowe1967,Glorieux2012}:
\begin{equation}
\frac{\partial}{\partial t}\rho_{01}(\mathbf{r},t)=D\nabla^{2}\rho_{01}(\mathbf{r},t)-\gamma_{0}\rho_{01}(\mathbf{r},t).\label{eq:diff}
\end{equation}
This equation can be readily solved in Fourier domain:
\begin{equation}
\rho_{01}(\mathbf{r},t)=e^{-\gamma_{0}t}\sum_{\mathbf{K}}\beta(K)e^{-DK^{2}t}e^{i\mathbf{K\cdot r}}.\label{eq:coherencedecay}
\end{equation}

Evolution of each plane wave component of initial amplitude $\beta(K)$
and wave vector $K$ is described by a simple exponential decay at
a rate $\gamma(\mathbf{K})=\gamma_{0}+DK^{2}$. As long as the evolution
of $\rho_{01}$ can be described by Eq. \eqref{eq:diff} with position-independent
homogeneous depolarization with a rate $\gamma_{0}$, the measurement
of decay rates $\gamma(K)$ is sufficient to calculate $D$ as a coefficient
of the quadratic term of $\gamma(K)$.

\subsection{Creation and probing of atomic coherence}

\begin{figure}
\includegraphics[width=8cm]{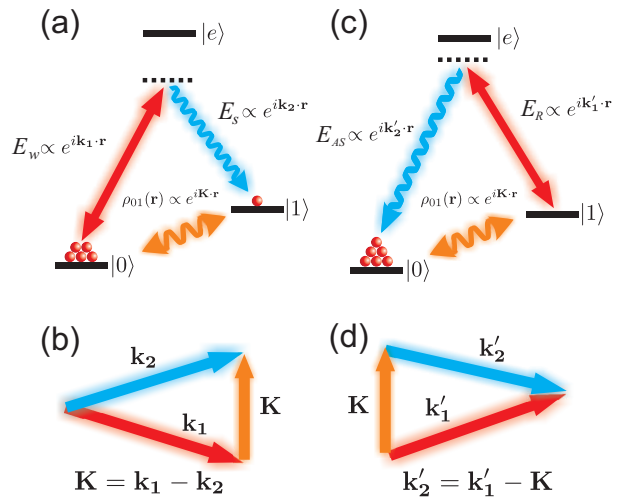} \centering \caption{(Color online) (a) Atomic coherence $\rho_{01}(\mathbf{r})$ is created
in spontaneous Stokes scattering. (b) The difference between wave
vectors of laser field \textbf{$\mathbf{k_{1}}$} and the scattering
field component \textbf{$\mathbf{k_{2}}$} is stored in the spatially
dependent phase of the atomic coherence. (c) Atomic coherence can
be converted back to light in anti-Stokes scattering. (d) The atomic
coherence phase is imprinted on anti-Stokes scattering as momentum
conservation. Unlike the Stokes process, anti-Stokes scattering deterministically
retrieves created coherence pattern on demand. \label{fig:Levels}}
\end{figure}

In Fig. \ref{fig:Levels}(a) we illustrated the atomic levels involved
in Stokes scattering leading to creation of coherence $\rho_{01}(\mathbf{r})$
between levels $\ket0$ and $\ket1$. Upon spontaneous scattering
of a laser beam detuned from the $|0\rangle\leftrightarrow|e\rangle$
transition both scattered light and atomic coherence are created.
We shall consider a simple case where the laser beam and the scattered
light are planewaves with wave vectors, \textbf{$\mathbf{k_{1}}$}
and \textbf{$\mathbf{k_{2}}$}, respectively. The difference between
the laser field wave vector $\mathbf{k_{1}}$ and the created photons
wave vector $\mathbf{k_{2}}$ is accumulated in atoms as a spatial
phase of atomic coherence. As illustrated in Fig. \ref{fig:Levels}(b)
the atomic coherence created will be of a form $\rho_{01}(\mathbf{r})=\beta e^{i\mathbf{K\cdot r}}$,
where \textbf{$\mathbf{K=k_{1}-k_{2}}$}.

Such periodic atomic coherence $\rho_{01}(\mathbf{r})$ can work as
a diffraction grating and deflect a laser beam. This is realized in
anti-Stokes scattering process presented in Fig. \ref{fig:Levels}(c)
in which laser beam detuned from the $|1\rangle\leftrightarrow|e\rangle$
transition is scattered at an angle. The spatial phase of atomic coherence
$\rho_{01}(\mathbf{r})$ is imprinted back onto the scattered photons
as follows from Bragg condition. Provided driving laser beam is the
plane wave with a wave vector $\mathbf{k_{1}'}$, the diffracted light
wave vector will be $\mathbf{k_{2}'=k_{1}'-K}$, as illustrated in
Fig. \ref{fig:Levels}(d). Therefore by observing the intensity of
the light $I_{\mathrm{AS}}(\theta)$ scattered at an angle $\theta$,
we register a signal which is proportional to the modulus square of
the corresponding plane wave component of atomic coherence $|\beta(K)\exp(-\gamma(K)t)|^{2}$,
with $\theta=K/k_{1}'=K\lambda/2\pi$.

\subsection{Averaging and retrieving the diffusion coefficient }

Spatially varying atomic coherence $\rho_{01}(\mathbf{r})$ is created
in a spontaneous Stokes scattering process, which populates various
planewave components randomly. Nonetheless, the average modulus square
of the excitation amplitude $\langle|\beta(K)|^{2}\rangle$ created
right after the scattering is set by the driving pulse parameters
and can be kept constant between measurement series. Therefore, we
can calculate the average intensity of the light scattered at a certain
angle $\theta=K/k_{1}'$ and for a given storage time $\tau$, incorporating
Eq. \eqref{eq:coherencedecay}:

\begin{equation}
\langle I_{\mathrm{AS}}(\theta=K/k_{1}',\tau)\rangle=\eta(K)\langle|\beta(K)|^{2}\rangle e^{-2\gamma(K)\tau},\label{eq:ASamp}
\end{equation}
where $\eta(K)$ is efficiency of readout, \emph{a priori} dependent
on $K$. The only factor that depends on the diffusion time $\tau$
is the intensity decay factor $e^{-2\gamma(K)\tau}$, which provides
direct information about the decay rate $\gamma(K)$. Therefore for
a given angle of observation $\theta=K/k_{1}'$ we can infer $\gamma(K)$
from an exponential fit to a series of experimental data taken for
successive $\tau$.

By repeating the decay fits for many $K$ we can gather and then fit
the expected functional dependence $\gamma(K)=\gamma_{0}+DK^{2},$
to obtain the diffusional coefficient $D$. In principle, we could
use only two measurements of decay rates $\gamma$ corresponding to
just two different directions. Note that thanks to a particular way
of populating spatially varying atomic coherence we create and probe
many wave vectors concurrently without altering the setup. Thence
we obtain many points corresponding to a broad span of $K$ vectors
in a single measurement sequence, which provides for a robust quadratic
fit of $\gamma(K)$. The quality and reliability of the experimental
data is directly reflected in this last fit.

\subsection{Angular blurring at readout}

So far we have assumed the driving beam to be a planewave. The finite
size $w$ of the driving laser beam in anti-Stokes scattering results
in limited resolution in probing a wave vector space. The angular
spread of the laser beam driving the readout will be transferred onto
the angular distribution of the scattered light due to momentum conservation
even for planewave atomic coherence $\rho_{01}(\mathbf{r})$. Thus
for any specific angle of observation $\theta=K/k_{1}'$ we detect
the scattered light originating from several distinct Fourier components
of the atomic coherence pattern. The contribution will come from the
component of the wave vector $K$ and its vicinity to the spread $\sigma$.
We expect the spread $\sigma$ to be of the order of the inverse of
the driving laser beam size $1/w$.

The result of the limited resolution is an overall increase in the
decay rates observed $\gamma_{\mathrm{obs}}$. It can be estimated
by convolving the storage time-dependent Fourier distribution of atomic
coherence $\rho_{01}(\mathbf{r})$ with a Gaussian of a spread $\sigma$,
yielding:

\begin{equation}
\gamma_{\mathrm{obs}}(K)=(\gamma_{0}+2D\sigma^{2})+DK^{2}.\label{eq:gobs}
\end{equation}

Note that the term quadratic in $K$ in the above formula did not
change, therefore the procedure of obtaining the diffusion coefficient
$D$ remains unchanged. We only have to assure that the term $2D\sigma^{2}$
is small as compared to $DK^{2}$. This can be done by increasing
the beam size $w$.

In conclusion all measurements can be completed by varying only one
parameter -- the diffusion time $\tau$ equal to laser pulse separation
while collecting scattered light on a camera. Data analysis require
three straightforward steps: averaging, exponential fit and eventually
quadratic fit to obtain $D$. This makes the whole procedure relatively
quick and simple to repeat.

\subsection{Additional sources of decoherence}

In Eq. \eqref{eq:diff} we assumed that decoherence could be divided
into two types: the $K$-dependent diffusional  type  and the homogenous
type. The latter originates mostly from atomic collisions. This division
applies sufficiently to typical experimental conditions; however, other
types of processes may contribute to decoherence as well.

In Eq. \eqref{eq:diff} we neglected the stray magnetic field. Taking
such a field into account would require introducing an extra term
in Eq. \eqref{eq:diff}: $i\mu_{B}(g_{1}m_{1}-g_{0}m_{0})B(\mathbf{r})\rho_{01}(\mathbf{r},t)$
to represent additional, space-dependent build-up of the phase, where
$\mu_{B}g_{i}m_{i}$ is the magnetic moment for the $i$-th level.
Nevertheless, we estimated that our final results would change only
by  10\% in the presence of a magnetic field gradient of 0.5 Gauss/mm
or with a quadratically changing magnetic field of 10 mGauss/mm$^{2}.$
These values are considerably higher than in normal experimental conditions
even without applying magnetic shielding. 

We also noted that decoherence due to spin-exchange collisions might
lead to quite complicated effects if $\rho_{01}(\mathbf{r})$ was
a fast varying function of position. These collisions could alter
the state of the atoms at a different rate and at different points
in space, leading to a nontrivial space dependence of $\gamma_{0}$.
We calculated the rate of these collisions to be of the order of 1.5
kHz, which is negligible as compared to the diffusional decay caused
by the atoms leaving the laser beams.

\section{Experiment}

\begin{figure}
\includegraphics[width=8cm]{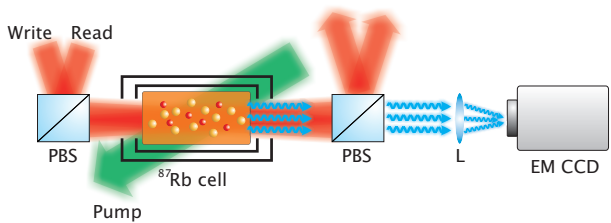} \centering \caption{(Color online) Central part of the experimental setup. Rubidium-87
atoms are mixed with the buffer gas in a glass cell inside a double
magnetic shielding. Pump, write and read are the laser beams used
to prepare atoms in the ground state $\ket0$, to create and probe
atomic coherence patterns respectively. Stokes and anti-Stokes scattering
are singled out on a polarizing beam splitter PBS and observed in
the far field on an electron multiplying EM CCD camera. \label{fig:setup}}
\end{figure}

In our case the levels $\ket0$ and $\ket1$ between which we create
atomic coherence $\rho_{01}(r)$ are hyperfine split levels $5_{2}\mathrm{S}_{1/2}$,
F=1 and F=2 respectively. Spatially dependent coherence $\rho_{01}(r)$
is created and read with the use of Raman transitions between this
levels. The scattered light is separated from much stronger stimulating
lasers by a polarizer and additionally filtered out spectrally by
a rubidium-85 filtering cell. The scattered light is registered by
an electron multiplying CCD (Hamamatsu) camera sensor which is placed
in the far field. The main part of the experimental setup is shown
in Fig. \ref{fig:setup}, further details can be found in \cite{Chrapkiewicz2012}.

\begin{figure}
\includegraphics[width=8cm]{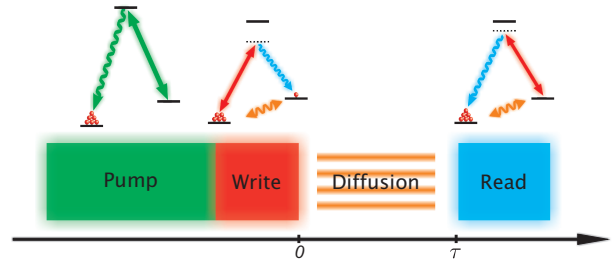} \centering \caption{(Color online) Single operational sequence. First rubidium atoms are
pumped on D2 line to F=1 ground level $\ket0$. Then spontaneous Stokes
scattering is driven to create spatially varying atomic coherence.
Next the coherence is stored in the dark for a time $\tau$ and we
fades due to diffusion. Finally the coherence driving anti-Stokes
scattering is probed. \label{fig:Sequence}}
\end{figure}

The operational sequence is shown in Fig. \ref{fig:Sequence}. Laser
pulses were formed with the use of acousto-optical modulators. We
began with optical pumping of rubidium atoms to the ground state F=1.
The pump laser operates in resonance on the D2 line. Then we created
random patterns of atomic coherence in the Stokes scattering process
driven by 2-5 $\mu$s pulses of the write laser detuned from the F
= 1$\rightarrow$F' = 1 D1 transition line by 1 GHz to the red. The
Stokes scattering was recorded with a camera for diagnostic purposes.
Next the atoms were left to diffuse for time $\tau$. Finally we used
5 $\mu$s long pulses of read laser detuned 1 GHz from the F = 2$\rightarrow$F'
= 2 D1 transition line to the blue to probe blurred atomic coherence
$\rho_{01}(\mathbf{r})$ and record anti-Stokes scattering intensity
in the far field $I_{\mathrm{AS}}(\theta)$.

Beam diameters $1/e^{2}$ and powers for the Stokes and anti-Stokes
drive laser were 5 mm, 4 mm and 16 mW, 7 mW respectively. They were
chosen in order to achieve both good resolution as discussed above
Eq. \eqref{eq:gobs} and a sufficient scattering rate \cite{Chrapkiewicz2012}.

We repeated the create-store-read sequence multiple times and recorded
random patterns of anti-Stokes scattering, changing the storage time
$\tau$.

We used four rubidium-87 cells with different buffer gases. Those
were: neon at the pressure of 2 Torr, krypton at 0.5 Torr and 1 Torr
and xenon at 1 Torr. All cells were 10 cm long, 2.5 cm diameter cylinders
made by Precision glassblowing. The longest times for anti-Stokes
scattering observations were ca. $\tau=$50 $\mu$s, which corresponded
to an RMS atomic displacement of about 1 mm, far less than the cell
size.

The cells temperature was stabilized at ca. 70$^{\circ}$ C, which
corresponded to an optical depth of 40 and a concentration of rubidium
atoms $n=10^{12}\ \mathrm{cm^{-3}}$. Cells were heated with bifilar
windings, but the heating current was interrupted for the time of
impulse sequence inducing Raman scattering. The cells were placed
inside a double magnetic shield. In the independent measurements we checked
if the quality of our shielding is good and we estimated the decoherence
rate due to imperfect shielding to be less than 1 kHz.

For each diffusion time $\tau$ we averaged 500 images of anti-Stokes
scattering each time obtaining smooth symmetric profiles. We subtracted
the averaged background. Given that the most important thing for us
is the intensity as a function of azimuth angle $\theta$, we also
carried out a polar averaging around the laser beam direction increasing
signal to noise ratio. The result of the measurements was average
scattering intensity $\langle I_{\mathrm{AS}}(\theta,\tau)\rangle$
as a function of angle $\theta$ and storage time $\tau$.

\section{Results}

\label{sec:results}

\begin{figure}
\includegraphics[width=8cm]{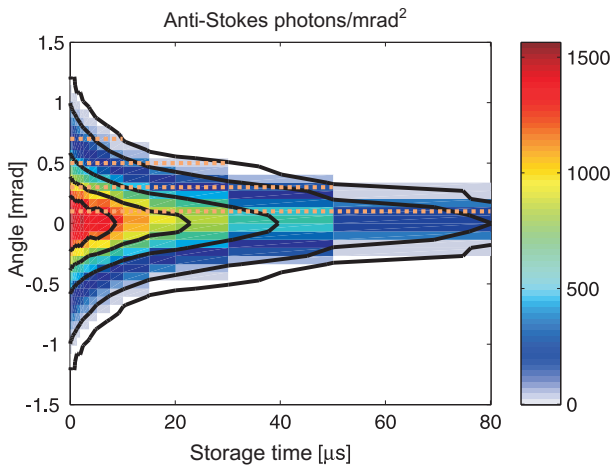} \centering \caption{(Color online) Average number of photons per pixel $\langle I_{\mathrm{AS}}(\theta,\tau)\rangle$
in the anti-Stokes scattering process as a function of the storage
time $\tau$ and angle of observation $\theta$. Data taken for a
cell with 1 Torr xenon. The decay is conspicuously faster for higher
angles of scattering. The contour lines show 1200, 800, 400, 50, 10
photons per pixel.\label{fig:ASMap} }
\end{figure}

In Fig. \ref{fig:ASMap} we present a typical map of the averaged
angular profiles of anti-Stokes scattering in cell with xenon vs free
depolarization time $\langle I_{\mathrm{AS}}(\theta,\tau)\rangle$. The results are given as a function of the scattering angle $\theta$,
which is proportional to the wave vector of the corresponding Fourier
components of atomic coherence patterns $\theta=K\lambda/2\pi$.

\begin{figure}
\includegraphics[width=8cm]{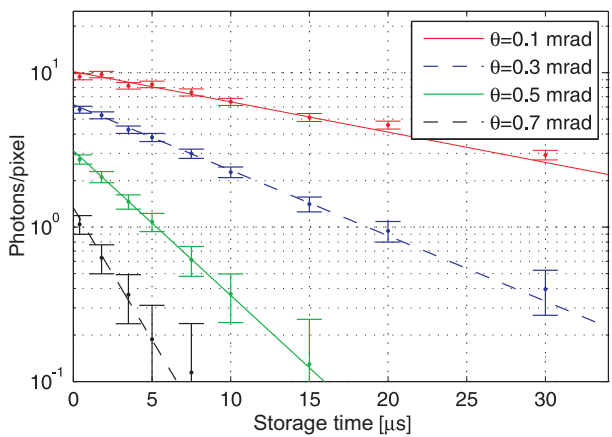} \centering \caption{(Color online) Average number of photons per pixel $\langle I_{\mathrm{AS}}(\theta,\tau)\rangle$
as a function of the storage time $\tau$ with exponential decay fits
observed for four different angles $\theta$. The decays are visibly
faster for higher scattering angles. The data corresponds to horizontal
dashed lines in Fig. \ref{fig:ASMap}. \label{fig:ASDecay}}
\end{figure}

In Fig. \ref{fig:ASDecay} we plot an average number of photons as
a function of the storage time $\tau$ at a few angles $\theta$ marked
with dashed lines in Fig. \ref{fig:ASMap}. The decay rate is faster
for higher angles of scattering. The error bars in Fig. \ref{fig:ASDecay}
correspond to 1$\sigma$ uncertainty and were calculated from the
full statistics of camera counts. We fit exponential decays to the
data taking the error bars into account. It is worth underlining that
the data fits well to the curve at each scattering angle. As shown
in Fig. \ref{fig:ASDecay}, for high scattering angles $\theta$ the
average signal $\langle I_{\mathrm{AS}}(\theta,\tau)\rangle$ is at
the level of one photon per shot which considerably increases the
uncertainty of $\gamma$ .

\subsection{Diffusion coefficients}

\begin{figure*}
\includegraphics[width=15cm]{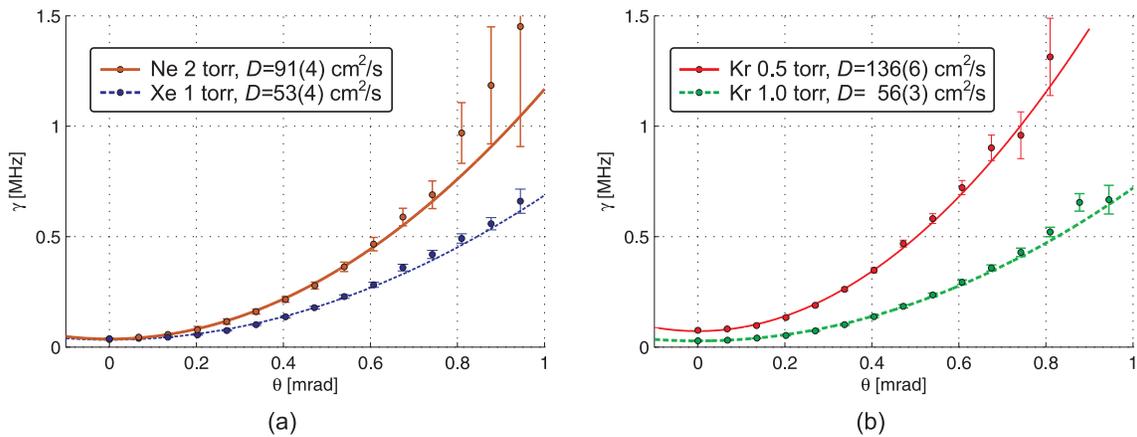} \centering \caption{(Color online) Decoherence rates $\gamma(\theta)$ as a function of
the scattering angle corresponding to the atomic coherence wave vectors
$K=2\pi\theta/\lambda$. We fit the data with quadratic dependence
$\gamma_{\mathrm{obs}}(\theta)=\gamma_{0,\mathrm{obs}}+D\theta^{2}(2\pi/\lambda)^{2}$.
Quadratic term of fit gives the diffusion coefficient $D$. (a) Data
taken for 1 Torr of xenon and 2 Torr of neon (b) Data taken for krypton
of different pressures 0.5 Torr and 1 Torr. \label{fig:XeNeKr}}
\end{figure*}

Having obtained an exponential fits for decays in all directions,
we can analyze decays rates $\gamma_{\mathrm{obs}}(\theta)$ as a
function of the deflection angle $\theta$. In Fig. \ref{fig:XeNeKr}(a)
we present fitted decay rates for measurements in cells filled with
1 Torr of xenon as well as with 2 Torr of neon . In Fig. \ref{fig:XeNeKr}
(b) we give measured data for cells with krypton at two pressure values:
0.5 Torr and 1 Torr.

As expected, the data fits to the quadratic dependence $\gamma_{\mathrm{obs}}(\theta)=\gamma_{0,\mathrm{obs}}+D\theta^{2}(2\pi/\lambda)^{2}$.
Now our $1\sigma$ error bars correspond to respective confidence
bound from the exponential fit. Note that for high decay rate values
the uncertainties are large because they correspond to small and noisy
signals.

The measurements described above were repeated several times in order
to make sure the results were reliable and repeatable. We changed
the amount of the light generated by altering pulse duration of the
writing laser and by repeating measurements at different temperatures.
The spread of diffusion coefficients obtained was about 5-12\%, depending
on the cell. We attribute this spread to the beam wander, laser power
and frequency instability, and the drift of the temperature of the
cell during each measurement sequence.

\begin{table}[H]
\centering%
\begin{tabular}{ccrclc}
\hline 
Buffer gas  & Pressure {[}Torr{]}  & \multicolumn{3}{c}{$D\mathrm{[cm^{2}}/\mathrm{s}]$} & $\gamma_{0,\mathrm{obs}}$ {[}kHz{]}\tabularnewline
\hline 
Ne  & 2  & 91  & $\pm$  & 11  & 38\tabularnewline
Kr  & 0.5  & 136  & $\pm$  & 9  & 71\tabularnewline
Kr  & 1  & 57.5  & $\pm$  & 3  & 28\tabularnewline
Xe  & 1  & 52  & $\pm$  & 3  & 33\tabularnewline
\hline 
\end{tabular}

\caption{Fit parameters for measured cells. Temperature: 70$^{\circ}$ C .
\label{tab:Dgamma}}
\end{table}

In Tab. \ref{tab:Dgamma} we summarize fitted values from the measured
data from charts in Fig. \ref{fig:XeNeKr}. Note, that for measurements
in krypton the ratio of the obtained diffusion coefficients is close
to the nominal pressure values.

To further verify the accuracy of our results, we carried out reference
measurements for krypton at 1 Torr using write and read beams reduced
ca. 3 times, so that their diameters $1/e^{2}$ were 1.6 mm i 1.4
mm respectively. This time the measurement was definitely less accurate
due to the spread of wave vectors of the read beam and due to aberrations
in the imaging system. The diffusion coefficient measured lay within
the range of $40\ \mathrm{cm^{2}/s}$ to $65\ \mathrm{cm^{2}/s}$,
which is consistent with other results .

Finally let us note, that the observed decay rate at $K=0$, $\gamma_{0,\mathrm{obs}}$
summarized in Tab. \ref{tab:Dgamma} is dominated by excessive contribution
due to the finite read beam size, $2D\sigma^{2}$ in Eq. \eqref{eq:gobs}.
From respective collisional cross sections \cite{Franz1965} we estimate
ca. $1.5$ kHz of decay rate due to Rb-Rb collisions and contributions
less than 150 Hz from Rb -- buffer gas collisions.

\subsection{Normalized diffusion coefficients}

\begin{table*}[!t]
\centering%
\begin{tabular}{ccc}
\hline 
Buffer gas  & $D_{0}\mathrm{[cm^{2}}/\mathrm{s}]$ -- this paper  & $D_{0}\mathrm{[cm^{2}}/\mathrm{s}]$ - previous results \tabularnewline
\hline 
Ne  & 0.20$\pm$0.02  & 0.11 \cite{Shuker2008}, 0.18 \cite{Vanier1974}, 0.31 \cite{Franzen1959,Arditi1964},
0.48 \cite{Franz1965}\tabularnewline
Kr  & 0.068$\pm0.006$  & 0.1 \cite{Bouchiat1972} 0.04 \cite{Higginbottom2012}\tabularnewline
Xe  & 0.057$\pm$0.007  & No experimental data\tabularnewline
\hline 
\end{tabular}

\caption{Measured diffusion coefficients of rubidium atoms in noble buffer
gases: Ne, Kr, Xe. The results are normalized to 0$^{\circ}$C and
760 Torr pressure. \label{tab:NormalizedD}}
\end{table*}

To compare our results with literature, we normalized the results
to the standard conditions, i.e. at atmospheric pressure and at 0$^{\circ}$
C using the standard approximate formula \cite{Happer1972}, however
strict scaling in a wide range of temperatures and pressures would
involve Chapman-Enskog theory \cite{Chapman1970}

\begin{equation}
D_{0}=D\Big(\frac{P}{760\ \textrm{Torr}}\Big)\sqrt{\frac{T_{0}}{T}},
\end{equation}
where $P$ stands for the gas pressure at $T_{0}=0{}^{\circ}$C, and
$T$ is the temperature upon measurement. The diffusion coefficient
will scale as $T^{1/2}$ according to \cite{Hogervorst1971}.

The temperature is known with good accuracy and the main error in
determining $D_{0}$ results from inaccuracy of gas pressure in cells,
specified by the manufacturer not to be worse than 10\%, and from
the spread of the measured $D$ values. The normalized diffusion coefficients
are listed in Tab. \ref{tab:NormalizedD} together with the data published
previously.

The results we obtained do not differ from those obtained beforehand
which -- as we can see from Tab. \ref{tab:NormalizedD}-- were characterized
by a noticeable spread.

\section{Conclusions}

We have demonstrated a novel method for measuring diffusion coefficients
of atoms tailored to atomic memory applications. It should be emphasized
that the method allows for singling out contribution of the diffusion
in any given cell without prior knowledge of other decoherence mechanisms.
It is based on creation of spatially varying atomic coherence fields,
letting them diffuse in the dark and probing them. The fields comprise
various spatial periodicities evolving concurrently, created and probed
with Raman scattering. Due to the diffusion distinct components decay
at different rates. All other significant sources of decoherence are
homogenous and contribute equally to the decay of all components.
Therefore we can extract diffusion coefficients by measuring decay
rates for different periodicities of different components of spatially
varying atomic coherence. Distinct components are conveniently mapped
on different angles of the scattered light which enables observations
with a camera in the far field. 

Our method does not require any setup alterations within the measurement.
This leads to quite direct determination of the diffusion coefficient
founded on a basic time and angle calibrations of the experimental
setup. 

We have made sure that the method is accurate and repeatable. The
results are based on multiple independent measurements for a number
of various periodicities which give almost the same values. We also
checked that varying laser beam widths and detunings does not affect
the final result and that it scales properly with gas pressure.

We suppose that our method could be incorporated into experiments
in which diffusion is the limiting factor, such as EIT, quantum memories
including gradient echo memory or collective Raman scattering, by
relatively straightforward modifications, such as adding a pump laser
and a camera in the far field. Other technical requirements are typically
fulfilled since these experiments also rely on using lasers of a few
MHz frequency stability and magnetic shielding.

We have measured diffusion coefficients of rubidium in neon, krypton
and xenon. Reliable values of diffusion coefficients in these gases
facilitate setup design and data interpretation in the multimode quantum
storage experiments. Moreover we recommend the use of xenon as a buffer
gas in case of Raman interaction and, to the best of our knowledge,
we provide the first experimental value of diffusion coefficient in
this gas. We believe that this value can also help develop experiments
with hyperpolarized xenon.

\section{Acknowledgments}

We acknowledge the generous support from Konrad Banaszek and Rafa\l{}
Demkowicz-Dobrza\'{n}ski. This work was supported by the Foundation
for Polish Science TEAM project, EU European Regional Development
Fund and FP7 FET project Q-ESSENCE (Contract No. 248095), National
Science Centre grant no. DEC-2011/03/D/ST2/01941 and by Polish NCBiR
under the ERA-NET CHIST-ERA project QUASAR.

%

\end{document}